\documentclass[twocolumn,showpacs,preprintnumbers,amsmath,amssymb,superscriptaddress]{revtex4}

\usepackage{graphicx}
\usepackage{dcolumn}
\usepackage{bm}

\def \ecm{$e\,$cm}

\def \upperlimit{$\left| d_{n}\right| <2.9 \times 10^{-26}$ \ecm}

\def \dBzdz{\partial B_z/\partial z}

\begin{document}

\title{Reply to Comment on ``An Improved Experimental Limit on the Electric Dipole Moment of the Neutron''
}
\author{C.A. Baker}\affiliation{Rutherford Appleton Laboratory, Chilton, Didcot, Oxon OX11 0QX, UK}
\author{D.D. Doyle}\affiliation{Department of Physics and Astronomy, University of Sussex, Falmer, Brighton BN1 9QH, UK}
\author{P. Geltenbort}\affiliation{Institut Laue-Langevin, BP 156, F-38042 Grenoble Cedex 9, France}
\author{K. Green}\affiliation{Rutherford Appleton Laboratory, Chilton, Didcot, Oxon OX11 0QX, UK}
\affiliation{Department of Physics and Astronomy, University of Sussex, Falmer, Brighton BN1 9QH, UK}
\author{M.G.D. \surname{van der Grinten}}\affiliation{Rutherford Appleton Laboratory, Chilton, Didcot, Oxon OX11 0QX, UK}\affiliation{Department of Physics and Astronomy, University of Sussex, Falmer, Brighton BN1 9QH, UK}
\author{P.G. Harris}\affiliation{Department of Physics and Astronomy, University of Sussex, Falmer, Brighton BN1 9QH, UK}
\author{P. Iaydjiev\footnote{On leave from Institute of Nuclear Research and Nuclear Energy, Sofia, Bulgaria}}\affiliation{Rutherford Appleton Laboratory, Chilton, Didcot, Oxon OX11 0QX, UK}
\author{S.N. Ivanov\footnote{On leave from Petersburg Nuclear Physics Institute, Russia}}\affiliation{Rutherford Appleton Laboratory, Chilton, Didcot, Oxon OX11 0QX, UK}
\author{D.J.R. May}\affiliation{Department of Physics and Astronomy, University of Sussex, Falmer, Brighton BN1 9QH, UK}
\author{J.M. Pendlebury}\affiliation{Department of Physics and Astronomy, University of Sussex, Falmer, Brighton BN1 9QH, UK}
\author{J.D. Richardson}\affiliation{Department of Physics and Astronomy, University of Sussex, Falmer, Brighton BN1 9QH, UK}
\author{D. Shiers}\affiliation{Department of Physics and Astronomy, University of Sussex, Falmer, Brighton BN1 9QH, UK}
\author{K.F. Smith}\affiliation{Department of Physics and Astronomy, University of Sussex, Falmer, Brighton BN1 9QH, UK}

\date{\today}

\pacs{13.40.Em, 07.55.Ge, 11.30.Er, 14.20.Dh}
\maketitle

Our Letter \cite{baker06} places a new experimental limit on the electric dipole moment (EDM) of the neutron.  The Comment \cite{lamoreaux06} points out that we did not explicitly include in our analysis the effect of the Earth's rotation, which shifts all of the frequency ratio measurements $R_a$ to lower (higher) values by 1.3 ppm when the $B_0$ field is upwards (downwards).  
However, this effect is essentially indistinguishable from other effects that can shift $R_a$, and all such shifts were compensated for in \cite{baker06} by using experimentally determined values of $R_{a0}$ (which we call $R_{a0\downarrow}$ and $R_{a0\uparrow}$, respectively, for the two polarities of $B_0$), where $\langle \partial B_z /\partial z \rangle_V$ = 0.

We turn now to the details.  Naively, one would expect that the crossing point of the lines in Fig.~2 of \cite{baker06} (which lies at $R_a-1 = 5.9\pm0.8$ ppm) would have $\langle \partial B_z /\partial z \rangle_V = 0$, with its ordinate yielding the true EDM.  However, what were referred to in \cite{baker06} as horizontal quadrupole fields (involving $\partial B_x/\partial y$ etc.) shift these lines towards the right.  A difference in the strengths of these quadrupolar fields upon $B_0$ reversal leads to a differential shift in $R_a$, and thus  to a vertical displacement of the crossing point.  The Earth's rotation mimics this behavior precisely, by moving the $B_0$-down (-up) line leftwards (rightwards).   
Thus, where quadrupole fields are mentioned in \cite{baker06}, one might better read this as ``quadrupole fields and Earth-rotation effects combined''.    The  ``quadrupole shift'' listed in Table 1 of \cite{baker06} simply represents the move from the crossing point to the average of the EDM values determined (independently) by the measured $R_{a0\downarrow}$ and $R_{a0\uparrow}$ values.  

The shift measurements are described (rather than just ``mentioned'') in \cite{baker06}.  First, the strongest constraint arises from a study of the depolarization of the neutrons as a function of $R_a$, and thus, effectively, as a function of $\langle \partial B_z /\partial z \rangle_V$.  Neutrons of different energies have different heights of their centers of mass, and thus the $T_2$ spin relaxation is maximized when$\langle (\partial B_z /\partial z)^2 \rangle_V$ is minimized.  The values of $R_a-1$ at which the polarization product $\alpha$ was found to peak were $(5.7\pm0.2, 5.9\pm0.2)$ ppm for $B_0$ up, down respectively. In the presence of the dipole in the region of the door of the storage chamber \cite{baker06}, the point for $B_0$ down (up) at which $\langle (\partial B_z /\partial z)^2 \rangle_V$ is minimized is 0.2 ppm higher (lower) than the point $R_{a0\downarrow}$ ($R_{a0\uparrow}$).   These data provide direct, independent measurements for each $B_0$ polarity of the actual values $R_{a0}$ at which $\langle \partial B_z /\partial z \rangle_V = 0$, taking into account any and all shift mechanisms, known or unknown, acting on $R_a$.  Since these depolarization results are drawn from the EDM data themselves, they cannot be described as ``{\it ex post facto}''. We conclude from our data that the differential quadrupole shift and Earth rotation effect cancel to within 15\% in our apparatus.  The fact that the resulting $d_n$ values (($-0.6\pm2.3, -0.9\pm 2.3) \times 10^{-26}$ \ecm\ for $B_0$ up, down respectively) agree so well with each other gives added confidence in the experimental results overall.

  Second, after about 60\% of the data had been taken, a bottle of variable height was used to measure the profile of the magnetic field within the storage volume.  Extrapolation of these data to the EDM bottle (which does include a small correction due to Earth's rotation) yields  $R_{a0\uparrow}-R_{a0\downarrow} = (1.5\pm1.0)$ ppm.  

Our data show no evidence for changes in the relevant long-term $B$-field properties from the periodic disassembly of the magnetic shields.

Since the  publication of \cite{baker06}, we have improved our fitting procedure to take full account of correlations between the quadrupole and dipole corrections, and to include explicitly the effect of the  Earth's rotation.   The results yield new net shifts (to be compared with those listed in Table 1 of \cite{baker06}) for the dipole and combined quadrupole/Earth rotation effects of $(-0.46, +0.30) \times 10^{-26}$ \ecm\ respectively, with a net uncertainty of $0.37\times10^{-26}$ \ecm\ for both.   In combination with the other effects discussed in \cite{baker06}  this yields an overall systematic correction to the crossing point of $(0.20\pm0.76)\times10^{-26}$ \ecm\  for the second analysis of \cite{baker06}.  The final value for the EDM from this analysis is then  
$(-0.4\pm1.5{\rm (stat)}\pm0.8 {\rm (syst)}) \times10^{-26} \,e\,{\rm cm}, 
$
implying $|d_n|<2.8\times10^{-26}$ \ecm\ (90\% CL), identical to the previous limit from this analysis.

The Comment asserts incorrectly that the $R_a-1$ values averaged to zero in the first analysis of \cite{baker06}.  By choice of the applied $\dBzdz$, they averaged to 8.9 ppm for both $B_0$ polarities.  
Since any net differential shifts in $R_a$ have been shown to be small, this analysis need not be altered.

In conclusion, the overall limit of \upperlimit\ (90\% CL) remains unchanged.

\end{document}